\documentclass[12pt]{article}
\usepackage{graviton}

\title{Scalar-hairy AdS Black Hole in the Einstein-Maxwell-Scalar Theory: first-order phase transition with a critical point}
\author[1]{Hong Guo\email{guohong@ibs.re.kr}}
\author[2]{Hang Liu\email{hangliu@sjtu.edu.cn}}
\author[3]{Yun Soo Myung\email{ysmyung@inje.ac.kr}}
\affil[1]{Particle Theory and Cosmology Group, Center for Theoretical Physics of the Universe, Institute for Basic Science (IBS), Daejeon 34126, Republic of Korea}
\affil[2]{College of Physics and Materials Science, Tianjin Normal University, Tianjin 300387, China}
\affil[3]{Center for Quantum Spacetime, Sogang University, Seoul 04107, Republic of Korea}

\reportnumber{CTPU-PTC-25-42}

\abstract{
In asymptotically anti–de Sitter (AdS) spacetime, we consider a real massive scalar field in the Einstein–Maxwell–scalar (EMS) model and examine both scalar-hairy black hole solutions induced by the nonminimal coupling to the Maxwell field and tachyonic-hairy solutions driven by the scalar potential.
When the scalar potential vanishes, scalar-hairy black holes emerge with profiles and properties similar to those observed in flat spacetime.
The presence of the scalar potential additionally induces tachyonic-hairy solutions, leading to the coexistence of these two distinct hairy phases in different regions of the parameter space.
The phase diagram reveals a first-order phase transition line between the tachyonic-hairy and scalar-hairy phases, originating at a critical point in the extreme temperature and chemical potential regime.
Our detailed analysis shows that this phase transition is directly associated with the self-overlap region of the scalar-hairy phase and its start point.
Moreover, increasing the coupling strength $\lambda$ shifts the critical point to higher temperature and chemical potential. 
}

\begin{document}
\maketitle

\section{Introduction}\label{sec=intro}

While general relativity (GR) has achieved remarkable success in weak-field tests, its predictions in the strong-field regime are now being scrutinized through gravitational-wave observations. 
These include the detection of binary black hole mergers by ground-based interferometers~\cite{LIGOScientific:2014pky,Aso:2013eba}, as well as the upcoming studies of extreme mass-ratio inspirals (EMRIs)~\cite{Amaro-Seoane:2007osp,Babak:2017tow}, supermassive black hole mergers~\cite{Begelman:1980vb,Volonteri:2002vz}, and other gravitational-wave sources~\cite{Barack:2018yly} by space-based detectors~\cite{LISA:2017pwj,Hu:2017mde,TianQin:2015yph}. 
Research on strong-field gravity is closely tied to some of the most profound theoretical challenges in modern physics. 
For example, phenomena such as the accelerated expansion of the Universe~\cite{Peebles:2002gy,Weinberg:2013agg} and the quest for a consistent unification of general relativity and quantum mechanics~\cite{Carlip:2001wq} remain unresolved within the standard framework of GR. 
These open questions have inspired extensive efforts to extend or modify Einstein’s theory~\cite{Clifton:2011jh,Heisenberg:2018vsk}, typically by introducing additional degrees of freedom that push the boundaries of classical relativity. 
Consequently, strong-field gravitational tests have become a key arena for probing GR, constraining its possible extensions, and exploring fundamental theoretical issues~\cite{Maselli:2021men,Li:2021pxf,Barsanti:2022vvl,Wang:2023jah,Yagi:2016jml,Baibhav:2019rsa,Bian:2025ifp}.

Benefiting from the no-hair theorem~\cite{Kerr:1963ud,Chrusciel:2012jk}, the presence of extra “hair” in black holes, arising from new degrees of freedom in modified gravity theories, has become one of the primary signatures indicating deviations from GR in strong-field tests. 
Studies of hairy black holes, together with gravitational and electromagnetic signal observations, are therefore of great interest in theoretical directions of gravity.
Among possible fields, scalar fields are the simplest form of matter fields, playing fundamental roles in both GR and cosmology~\cite{Cardoso:2019rvt,Barbon:2009ya,Choptuik:1992jv,Marsh:2015xka}. 
When serving as additional black hole hair, they also provide the simplest realization of hairy black holes and have long served as an important theoretical framework for evading the no-hair theorem and exploring the mechanisms behind black hole hair formation~\cite{Herdeiro:2015waa,Volkov:2016ehx}.

Among various dynamical mechanisms that can give rise to hairy black holes, spontaneous scalarization has attracted considerable attention in recent years.
Analogous to the phenomenon of spontaneous magnetization in ferromagnets, the scalar field can spontaneously develop a nontrivial profile once certain model parameters exceed a critical threshold. 
This mechanism was first discovered in 1993 in the context of neutron stars. 
More recently, spontaneous scalarization of black holes was realized through the nonminimal coupling between the scalar field and the Gauss–Bonnet curvature~\cite{Doneva:2017bvd,Silva:2017uqg,Antoniou:2017acq}, which rapidly stimulated a surge of interest and further developments. 
Extensive research has since been carried out, including extensions to spacetimes with a cosmological constant~\cite{Bakopoulos:2018nui,Guo:2020sdu,Brihaye:2019gla}, spin-induced scalarization~\cite{Dima:2020yac,Herdeiro:2020wei}, and models where the geometric source term is replaced by matter invariants such as the Maxwell invariant~\cite{Herdeiro:2018wub,Zhang:2021nnn,Liu:2022fxy} or the Ricci scalar~\cite{Herdeiro:2019yjy}. 
A comprehensive overview and the recent progress in these topics can be found in the review~\cite{Doneva:2022ewd}.

The formation of hairy black holes through spontaneous scalarization is primarily understood in terms of tachyonic instability. 
On a fixed GR vacuum background, linear perturbations of the scalar field satisfy a massive Klein–Gordon equation
\begin{align}
	\left(\Box -\mu^2_{\text{eff}}\right)\delta\psi = 0,
\end{align}
where the effective mass term $\mu_{\text{eff}}^{2}$ arises from the nonminimal coupling between the scalar field and the source term. 
When this effective mass squared becomes negative in certain regions, scalar perturbations can grow exponentially, rendering the original vacuum black hole solution unstable~\cite{Perivolaropoulos:2020uqy}.
To trigger such tachyonic instability in Einstein–Maxwell–scalar (EMS) theory, the coupling function $f(\psi) = e^{\lambda\psi^2}$ is typically required to satisfy $d^{2}f/d\psi^{2} > 0$ and $df/d\psi < 0$, which correspond to a negative coupling constant $\lambda<0$~\cite{Herdeiro:2018wub,Fernandes:2019rez}. 
However, in our recent work~\cite{Guo:2025xwh}, we found that even in the absence of tachyonic instability, namely for $\lambda > 0$,the EMS theory can still admit an entirely new class of hairy black hole solutions distinct from scalarized black holes.

This newly discovered family of hairy black holes provides a novel configuration of black hole hair. 
Unlike conventional hairy black holes, in which the scalar field typically decays monotonically with the radial coordinate, the scalar field in this case increases monotonically with radius and approaches a finite constant value at spatial infinity~\cite{Guo:2025xwh}. 
It is therefore of particular interest to explore the physical properties of these scalar-hairy black holes in more detail. 
The distinctive, monotonically increasing scalar field profile may serve as an observable signature distinguishing such configurations from other coexisting hairy black hole solutions. 
Moreover, these differences could have important implications for the thermodynamic behavior and the possible existence of rich phase structures in the space of black hole solutions.

Motivated by these considerations, the main purpose of this work is to investigate scalar-hairy black hole solutions in asymptotically anti–de Sitter (AdS) spacetime and to analyze how the scalar field mass affects their structure. 
Our results demonstrate that, first, the EMS theory with a massless scalar field also admits scalar-hairy black holes in AdS spacetime that exhibit a monotonically increasing scalar profile. 
Second, the inclusion of a scalar mass term can induce tachyonic instability, giving rise to tachyonic-hairy black holes in AdS spacetime~\cite{Sudarsky:2002mk,Hertog:2004bb,Hertog:2004dr}.
Consequently, in the massive case, both tachyonic-hairy and scalar-hairy black holes can coexist.
In the dual boundary field theory, these two distinct hairy black hole phases occupy different regions of the phase space, and a first-order phase transition can be identified, which originates from a critical point. 
This phase transition reveals an equilibrium relationship between the two hairy phases, with the location of the phase transition line being tunable by varying the coupling parameter. 
Increasing the coupling strength shifts the critical point toward higher temperature and chemical potential.

It is worth noting that in our previous studies based on the simplified and improved Einstein–Maxwell–dilaton (EMD) models~\cite{Guo:2024ymo,Guo:2025rxh}, we also analyzed the existence of two distinct hairy black hole phases and uncovered a similarly rich phase structure. 
When a weak-field expansion is performed for both models, it becomes evident that, at leading order, the simplified and improved EMD models are equivalent to the EMS model in AdS spacetime with a massive scalar field for a positive coupling parameter. 
Building upon this correspondence, the present work provides a more rigorous and complete analysis, demonstrating that the complex phase structure and the observed first-order phase transition in the two EMD models originate from the interplay between the scalar-hairy black holes and the tachyonic-hairy black holes in AdS spacetime.

\section{Einstein-Maxwell-Scalar Model}\label{sec=model}

In this section, we introduce an Einstein-Maxwell-scalar theory in the 4-dimensional AdS spacetime, described by the action
\begin{equation}\label{eq=action}
	S=\int d x^4\sqrt{-g}\left(R-2\Lambda-\frac{1}{2} \nabla^\mu \psi \nabla_\mu \psi-\frac{1}{4} f(\psi) F_{\mu \nu}^2-V(\psi)\right),
\end{equation}
where a real massive scalar field $\psi$ with potential $V(\psi)$ is coupled to the Maxwell field tensor $F_{\mu\nu}$ through a coupling function $f(\psi)$.
The cosmological constant is given by $\Lambda=-\frac{3}{L^2}$, with $L$ denoting the AdS radius, and $\nabla_\mu$ represents the covariant derivative.

Variating the action~\eqref{eq=action} with respect to the metric, scalar and Maxwell fields yields the corresponding equations of motion
\begin{align}
   & \nabla_\mu \nabla^\mu \psi-\frac{\partial_\psi f}{4} F_{\mu \nu} F^{\mu \nu}-\partial_\psi V=0, \label{eq=scalar}\\
   & \nabla^\mu\left(f(\psi)F_{\mu\nu}\right)=0,\label{eq=maxwell}\\
   & R_{\mu \nu}-\frac{1}{2} R g_{\mu \nu}+\Lambda g_{\mu\nu}=\frac{1}{2}\left(T_{\mu\nu}^\psi+f(\psi)T_{\mu\nu}^{E}\right),\label{eq=metric}
\end{align}
where the scalar and electromagnetic parts of the energy-momentum tensors are given by
\begin{align}
   & T_{\mu\nu}^\psi = \nabla_\mu \psi \nabla_\nu \psi - g_{\mu\nu}\left(\frac{1}{2}\nabla_\rho\psi \nabla^\rho \psi + V(\psi)\right),\\
   & T_{\mu\nu}^E = F_{\mu\rho}F_\nu{ }^\rho - \frac{1}{4}g_{\mu\nu}F_{\sigma \rho}^2.
\end{align}

In this model, we consider a static and spherical symmetric spacetimes, where the hairy black hole solutions can be constructed through the following ansatz
\begin{align}
	ds^2 & = -g(r)e^{-\chi(r)}dt^2 +\frac{dr^2}{g(r)} + r^2 (d\theta^2+\sin^2\theta d\varphi^2),\\
	A & = A_t(r)dt,\\
	\psi & = \psi(r).
\end{align}
In this setup, the Maxwell field reduces to a purely electric configuration, and both of the electric and scalar fields depend only on the radial coordinate. 
The event horizon is located at $r_h$ where the metric function satisfies $g(r_h)=0$, while the asymptotic boundary corresponds to $r\rightarrow\infty$.

Under these conditions, the Hawking temperature and the horizon area can be obtained by
\begin{align}
	T_H=\frac{1}{4\pi}g'(r_h)e^{-\chi(r_h)/2},\quad A_H=4\pi r_h^2.
\end{align}
The corresponding field equations take the form
\begin{align}
	& \psi''(r) + \left(\frac{g'(r)}{g(r)}-\frac{\chi'(r)}{2}+\frac{2}{r}\right)\psi'(r) + \frac{e^{\chi(r)}A_t'(r)^2}{2 g(r)}f'(\psi)-\frac{1}{g(r)}\partial_\psi V=0, \label{eq=motion1}\\
	& A_t''(r)+ \left(\frac{\chi '(r)}{2}+\frac{2}{r}+\frac{f'(\psi)}{f(\psi)}\psi'(r)\right)A_t'(r)=0, \label{eq=motion2}\\
	& \frac{g'(r)}{r} + g(r)\left(\frac{1}{r^2}+\frac{1}{4}\psi'(r)^2\right) + \frac{1}{4}f(\psi)e^{\chi(r)}A_t'(r)^2 + \frac{1}{2}V(\psi) - \frac{1}{r^2}-\frac{3}{L^2}=0, \label{eq=motion3}\\
	& 2 \chi '(r)+r \psi '(r)^2=0. \label{eq=motion4}
\end{align}

From Eq.~\eqref{eq=motion2}, one identifies the presence of a massless electric field, from which the conserved electric charge $Q$ can be expressed as
\begin{equation}\label{eq=charge}
	Q=-r^2f(\psi)e^{\chi/2}A_t'(r).
\end{equation}
When the scalar field vanishes, the spacetime reduces to an electric-vacuum configuration, namely the RN black hole.
In contrast, when the scalar field develops a nontrivial profile, Eq.~\eqref{eq=motion4} shows that the deviation from the RN solution is governed by the scalar field, signaling the existence of hairy black hole configurations.
Finally, to avoid the singular behavior of the metric function $g(r\to\infty)$ in the asymptotically AdS region, we redefine it as $g(r) = g_B(r)/(r^2 L^2)$ for numerical convenience.

In present work, we adopt a scalar potential of the form $V(\psi) = \frac{1}{2} m^2 \psi^2$ and a coupling function $f(\psi) = e^{-\lambda \psi^2}$, following Ref.~\cite{Guo:2025xwh}. 
Accordingly, the effective mass appearing in the Klein–Gordon equation~\eqref{eq=scalar} takes the form
\begin{align}\label{eq=mass}
	\mu^2_{\text{eff}}=\frac{1}{4}F_{\mu\nu}F^{\mu\nu}\partial_\psi^2f(\psi)+\partial^2_\psi V\big|_{\psi\to 0}=\lambda e^{\chi(r)}A_t'(r)^2+m^2.
\end{align}
For a massless scalar field, it is straightforward to see that a negative effective mass squared arises when $\lambda<0$, which triggers the onset of spontaneous scalarization in both asymptotically flat and asymptotically AdS spacetimes~\cite{Herdeiro:2018wub,Guo:2021zed,Promsiri:2023yda}.
Our recent work~\cite{Guo:2025xwh} has shown that for $\lambda > 0$, a novel scalar-hairy black hole solution exists and remains stable against linearized radial perturbations in asymptotically flat spacetime.
When the nonminimal coupling term vanishes, the hairy black hole solutions are instead supported by the tachyonic instability  associated with a massive scalar field in the asymptotically AdS spacetime~\cite{Sudarsky:2002mk,Hertog:2004bb,Hertog:2004dr}.
In this case, a negative scalar mass squared renders $\mu_{\text{eff}}^{2}$ negative and induces a tachyonic instability. 
On the other hand, a positive value of $\lambda$ or an increase in the black hole charge $Q$, counteracts this negative contribution to $\mu_{\text{eff}}^{2}$, thereby stabilizing the system.

The boundary condition near the horizon also provide insight into the competition between the nonminimal coupling and the scalar mass. For numerical convenience, we introduce a compact radial coordinate defined by $z \equiv r_h / r$.
In terms of this coordinate, the near-horizon behaviors of the scalar, Maxwell, and metric fields corresponding to Eqs.~\eqref{eq=motion1}–\eqref{eq=motion4} can be expanded around $z \rightarrow 1$ as
\begin{align}
	\psi(z) &= \psi_0+\psi_1(z-1)+\cdots ,\label{eq=bdhorizon1}\\ 
	A_t(z)  &= a_1(z-1)+ a_2(z-1)^2+\cdots ,\label{eq=bdhorizon2}\\ 
	g_B(z)    &= g_1(z-1)+ g_2(z-1)^2+\cdots , \label{eq=bdhorizon3}\\ 
	\chi(z) &= \chi_1(z-1)+\cdots . \label{eq=bdhorizon4}
\end{align}
The leading-order expansion coefficients are given by
\begin{align}
	\psi_1 =-\frac{2C_{h1}}{C_{h2}}, \quad \quad
	g_1 =\frac{1}{4}C_{h2}, \quad \quad
	\chi_1 =\frac{2C_{h1}^2}{C_{h2}^2},
\end{align}
where the quantities $C_{h1}$ and $C_{h2}$ are two independent quantities defined as
\begin{align}
	C_{h1} &= a_1^2f'(\psi_0)-2\psi_0 V'(\psi_0)=-2\psi_0\left(a_1^2\lambda e^{-\lambda\psi_0^2}+m^2\psi_0\right),\label{eq=Ch1}\\
	C_{h2} &= a_1^2f(\psi_0)+2\psi_0^2V'(\psi_0)-16=a_1^2e^{-\lambda\psi_0^2}+m^2\psi_0^3-16.\label{eq=Ch2}
\end{align}
While $C_{h2}$ typically retains its sign across a broad range of parameter space, particularly when the scalar mass squared satisfies $m^2<0$, Eq.~\eqref{eq=Ch1} indicates that the near-horizon behavior of the fields is governed by the competition between the scalar potential and the nonminimal coupling.
The increase of coupling tends to enhance cause the scalar field, whereas the negative mass squared drives the scalar field to decay as one moves from the horizon ($z=1$) toward spatial infinity ($z=0$).

For the boundary conditions at infinity, the field solutions approach the asymptotic AdS configuration, which requires
\begin{align}
    \psi(z\rightarrow 0) &=\Psi_0 z^{\Delta_{-}} + \Psi_1 z^{\Delta_{+}} + \cdots, \quad  
    & A_t(z\rightarrow 0)&= \mu - \rho z + \cdots,\\ 
	\chi(z\rightarrow 0) &= \chi_\infty^0 + \frac{1}{4}\Psi_0^2 z^{2\Delta_{-}} \cdots, \quad  
	& g_B(z\rightarrow 0)  &= 1 + L^2z^2 - 2 M L^2 z^3 + \cdots, 
\end{align}
where $M$ denotes the ADM mass of the black hole, while $\mu$ and $\rho$ correspond to the chemical potential and charge density in the dual field theory.
The coefficients $\Psi_0$ and $\Psi_1$ are integration constants associated with the scalar field.
The scaling dimensions are given by $\Delta_{\pm} = \frac{3 \pm \sqrt{9 + 4 m^2 L^2}}{2}$, and the Breitenlohner–Freedman (BF) bound requires the scalar mass to satisfy $m^2 > -\frac{9}{4L^2}$.

In particular, for $m^2 = 0$, the asymptotic behavior of the scalar field simplifies to
\begin{align}\label{eq=massless}
	\psi(z\rightarrow 0)=\Psi_0+\Psi_1 z^3.
\end{align}
The negativity of Eq.~\eqref{eq=Ch1} ensures that the scalar field grows from $z = 1$ to $z = 0$, approaching a nonvanishing constant value $\Psi_0$, as previously discussed in the asymptotically flat case~\cite{Guo:2025xwh}.
In the next section, we focus on the $m^2 = 0$ case to confirm the existence of scalar-hairy solutions in AdS spacetime.
Subsequently, we analyze the $m^2 = -2$ case, which yields the regular boundary condition for the scalar field
\begin{align}\label{eq=bdyinf}
	\psi(z\rightarrow 0)=\Psi_0 z+\Psi_1 z^2,
\end{align}
to explore the coexistence of two distinct hairy black hole branches and their corresponding phase transition in the boundary field theory.

Before proceeding, we note that the equations of motion admit three independent scaling symmetries
\begin{alignat}{3}
    t &\to \lambda_t t, &\qquad e^{\chi} &\to \lambda_t^2 e^{\chi}, &\qquad A_t &\to \lambda_t^{-1} A_t, \label{eq=scale1}\\
    r &\to \lambda_r r, &\qquad g        &\to \lambda_r^2 g,        &\qquad A_t &\to \lambda_r A_t,     \label{eq=scale2}\\
    r &\to \lambda_L r, &\qquad t        &\to \lambda_L t,          &\qquad L   &\to \lambda_L L.\label{eq=scale3}
\end{alignat}
where $\lambda_t$, $\lambda_r$ and $\lambda_L$ are dimensionless constants.
The first symmetry allows us to set $\chi(z\rightarrow 0)=\chi_\infty^0=0$, ensuring that the Hawking temperature matches the temperature of the boundary field theory in the context of the AdS/CFT correspondence.
The second and third symmetries permit us to normalize the horizon radius and the AdS radius to $r_h=1$ and $L=1$, respectively, which we adopt for numerical convenience in the subsequent calculations.

We solve the field equations~\eqref{eq=motion1}–\eqref{eq=motion4} numerically by integrating from the event horizon ($z=1$) outward to the boundary ($z=0$).
The numerical solutions are specified by the coupling parameter $\lambda$ and by the two horizon coefficients $(\psi_0,a_1)$.
To explore the main region of parameter space where hairy black hole solutions exist, we scan the scalar-hair parameter $\psi_0$ over the interval $[0.01,5]$.
The allowed range of $a_1$ is truncated by the condition $C_{h2}=0$ (Eq.~\eqref{eq=Ch2}), which signals the onset of a divergence in the near-horizon expansion.

\section{Scalar-hairy black hole with vanishing scalar mass}\label{sec=zeromass}

\begin{figure}[thbp]
    \centering
    \includegraphics[height=0.31\linewidth]{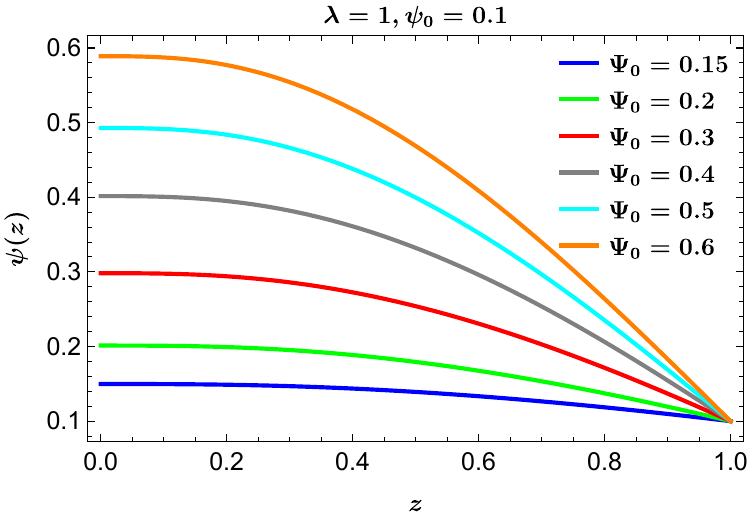}
    \includegraphics[height=0.31\linewidth]{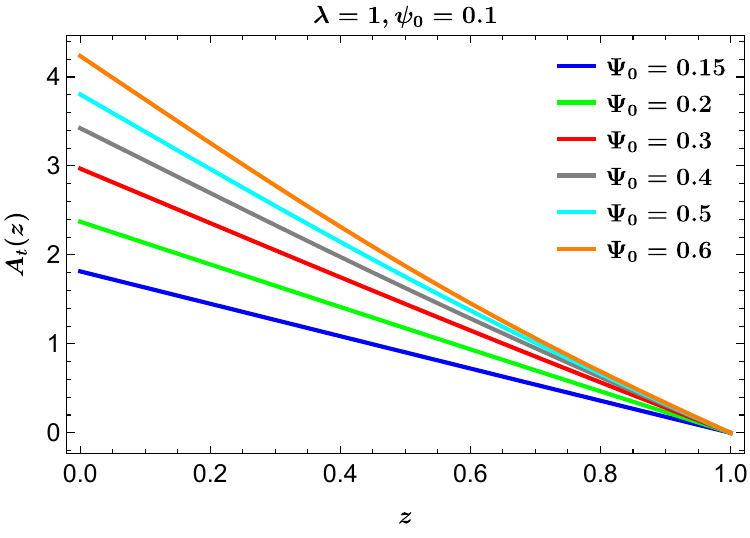}
    \includegraphics[height=0.31\linewidth]{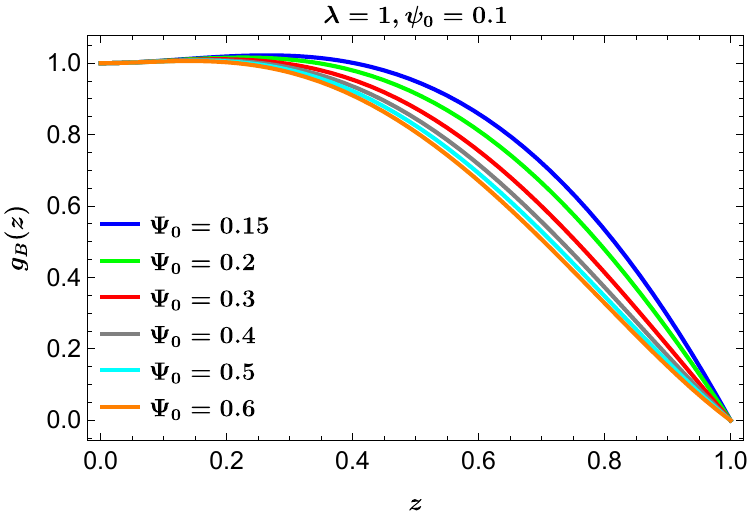}
    \includegraphics[height=0.31\linewidth]{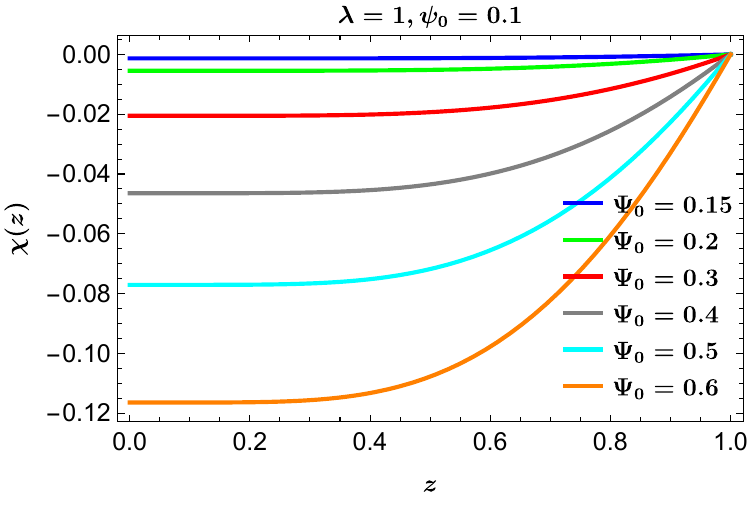}
    \caption{Profiles of the field functions for $\psi_0=1$ and $\lambda=1$, shown for different scalar charges $\Psi_0=0.15,0.2,0.3,0.4,0.5,0.6$.}
    \label{fig=profile_zero}
\end{figure}

In this section, we focus on the case of a massless scalar field ($m^2=0$) to confirm the existence of scalar-hairy black hole solutions in asymptotically AdS spacetime.
The profile of the scalar field is constrained by the near-horizon condition \eqref{eq=Ch1}, which requires the field to increase monotonically and approach a constant value in order to satisfy the boundary behavior \eqref{eq=massless} at infinity.
We employ the shooting method to obtain representative solutions corresponding to specific values of the scalar charge $\Psi_0$.
Fig.~\ref{fig=profile_zero} shows the profiles of the numerical field functions outside the event horizon for a fixed scalar hair $\psi_0=0.1$ and coupling parameter $\lambda=1$, with various shooting values of $\Psi_0$.
The scalar field profiles clearly demonstrate the existence of consistent solutions under different parameter choices.
The corresponding behaviors of the metric functions $g_B(z)$ and $\chi(z)$ reveal increasing deviations from the vacuum black hole as the scalar charge grows.
Moreover, the increase of $\Psi_0$ leads to a monotonic rise in the chemical potential, as seen from the Maxwell field profiles.
These features closely resemble those of scalar-hairy black hole solutions with increasing electric charge in asymptotically flat spacetime~\cite{Guo:2025xwh}, further confirming the existence of scalar-hairy black holes in the AdS background.

\begin{figure}[thbp]
    \centering
    \includegraphics[height=0.31\linewidth]{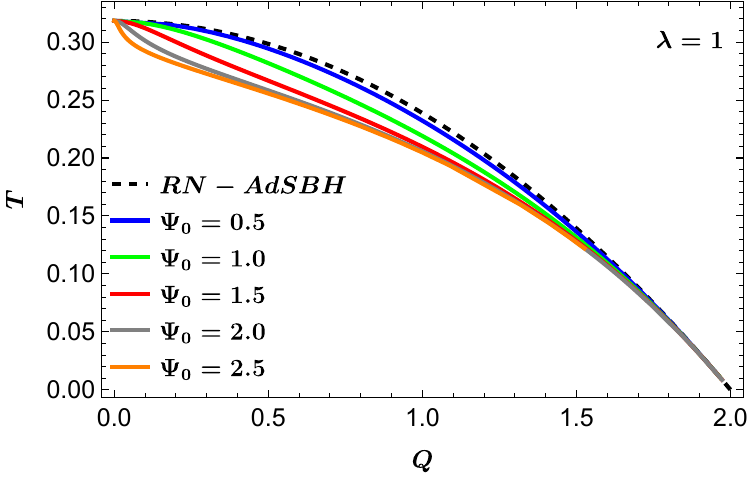}
    \includegraphics[height=0.31\linewidth]{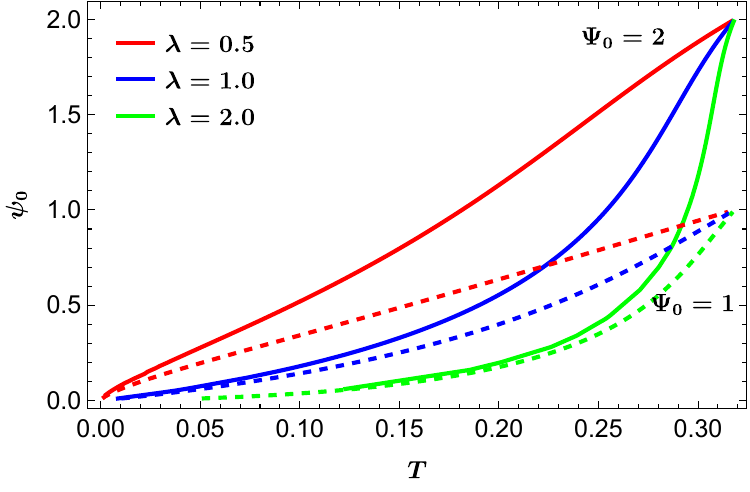}
    \caption{Left: Hawking temperature of the scalar-hairy black hole as a function of the electric charge $Q$ for different scalar charges $\Psi_0$; Right: Scalar hair as a function of $T$ for different coupling value $\lambda$ with $\Psi_0=1$ and $\Psi_0=2$, respectively.}
    \label{fig=temp_zero}
\end{figure}

To further investigate the thermodynamics of these hairy solutions, we plot the Hawking temperature as a function of the electric charge $Q$ in the left panel of Fig.~\ref{fig=temp_zero}.
The black dashed curve denotes the temperature–charge relation for the RN–AdS family.
As the scalar charge $\Psi_0$ increases, the solutions deviate progressively from the vacuum (RN–AdS) curve. 
The scalar-hairy black holes have consistently lower temperature than their RN–AdS counterpart, and larger $\Psi_0$ yields a lower temperature, in agreement with the asymptotically flat results of Ref.~\cite{Guo:2025xwh}.
A notable difference with the flat-space case is that here the deviation from the vacuum solution appears already in the small-charge region (left panel of Fig.~\ref{fig=temp_zero}), whereas in the flat case the deviation occurs only near the extremal charge.

Another interesting feature is that, as $\Psi_0$ increases, the branch of hairy solutions terminates before reaching the RN–AdS extremal charge $Q_{\mathrm{max}}=2$.
To clarify this point, the right panel of Fig.~\ref{fig=temp_zero} shows the scalar hair as a function of temperature for several values of the coupling $\lambda$.
We display results for two representative shooting values, $\Psi_0=1$ and $\Psi_0=2$.
When both $\lambda$ and $\Psi_0$ are small, the scalar hair grows continuously from zero as the temperature increases from zero, indicating a smooth onset of hairy solution from the extremal RN–AdS black hole.
By contrast, for larger coupling and scalar charge, the scalar hair exhibits a discontinuous jump from bald black hole when we increase the temperature to a critical value (green solid curve).
As discussed in the left panel, the increase of temperature corresponds to the process of discharge, so this jump can be explained by the discontinuous black hole transition from the large-charged RN-AdS black hole to the small-charged scalar-hairy black hole.

\begin{figure}[thbp]
    \centering
    \includegraphics[width=0.6\linewidth]{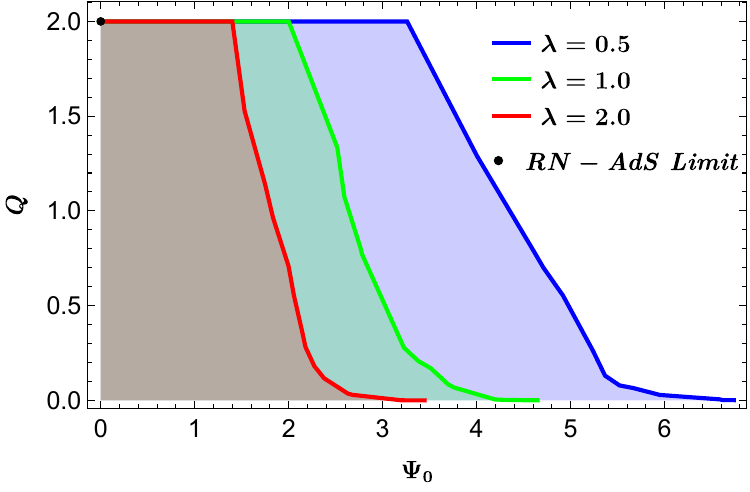}
    \caption{Existence domain of the scalar-hairy black holes in the parameter space of $(\Psi_0, Q)$. }
    \label{fig=phase_zero}
\end{figure}

The above results reveal the nontrivial thermodynamic features of scalar-hairy black holes and show that these solutions cease to exist when the system becomes overcharged.
We plot the domain of existence for scalar-hairy black holes in the parameter space spanned by the scalar charge $\Psi_0$ and the electric charge $Q$.
As illustrated in Fig.~\ref{fig=phase_zero}, the extremal charge $Q_{\text{max}}$ remains nearly constant for small scalar charge, consistent with the RN–AdS case.
However, once $\Psi_0$ exceeds a critical value $\Psi_0^{\text{cri}}$, the allowed charge range rapidly shrinks and eventually terminates at a maximum scalar charge $\Psi_0^{\text{max}}$.
Both $\Psi_0^{\text{cri}}$ and $\Psi_0^{\text{max}}$ decrease markedly as the coupling $\lambda$ increases.
For $\Psi_0 < \Psi_0^{\text{cri}}$, the continuous growth of the scalar hair from zero temperature indicates a smooth transition from the extremal RN–AdS black hole to a scalar-hairy configuration.
In contrast, when $\Psi_0 > \Psi_0^{\text{cri}}$, the shaded region in Fig.~\ref{fig=phase_zero} corresponds to scalar-hairy black holes with smaller electric charge, while the blank area above it represents RN–AdS black holes with larger charge.
This structure suggests a discontinuous transition between the non-extremal RN–AdS and scalar-hairy black hole phases.

\section{Scalar-hairy black hole with massive scalar field}\label{sec=negamass}

\subsection{Numerical solutions}\label{sec=negamass_profile}

We now turn to the case of a massive scalar field.
As noted above, scalar-hairy black holes in asymptotically AdS spacetimes can arise from the nonminimal coupling between the scalar and Maxwell fields.
Independently, a negative scalar mass squared can trigger a tachyonic instability and give rise to a distinct family of hairy solutions in AdS~\cite{Sudarsky:2002mk,Hertog:2004bb,Hertog:2004dr}.
From Eq.~\eqref{eq=Ch1}, one sees that when the nonminimal coupling term dominates near the horizon, the scalar tends to grow there; such a behavior typically produces a peak near the horizon followed by decay toward the boundary, consistent with the asymptotic form \eqref{eq=bdyinf}.
Conversely, when the scalar-potential term dominates, thereby triggering the tachyonic instability indicated by Eq.~\eqref{eq=mass}, the scalar decays monotonically and approaches zero at infinity.
For definiteness, we refer to the latter family as tachyonic-hairy black holes, while emphasizing that scalar hair need not always originate from a tachyonic instability.

By numerically integrating the field equations from the horizon to the boundary, we obtain hundreds of thousands of solutions across a wide region of the parameter space $(\psi_0,a_1)$.
In the dual field theory, $\Psi_0$ plays the role of the source for the scalar operator, setting an energy scale and breaking conformal symmetry in the boundary CFT~\cite{Cai:2022omk}.
To compare solutions on an equal footing, we fix this energy scale by holding $\Psi_0$ constant.
This is achieved by setting the reference scale $\Lambda_\psi=0.5$ and applying the scale transformation $r\to\lambda_\psi r$ with $\lambda_\psi=\Lambda_\psi/\Psi_0$ (cf. Eq.~\eqref{eq=scale2}).
So the scaled quantities transform as
\begin{align}
	\tilde{T}=\lambda_\psi T,\ \ \tilde{s}=\lambda_\psi^2 s, \ \ \tilde{\mu}=\lambda_\psi \mu,\ \ \tilde{\rho}=\lambda_\psi \rho,\ \ \tilde{\Psi}_1 =\lambda_\psi^2 \Psi_1, \ \ \tilde{Q}=Q.
\end{align} 
We note that trivial (scalar-free) RN solutions are sent to the edge of parameter space under this normalization and are therefore not considered in the following analysis; our focus is on nontrivial hairy black holes and their possible phase transitions.

\begin{figure}[thbp]
    \centering
    \includegraphics[height=0.31\linewidth]{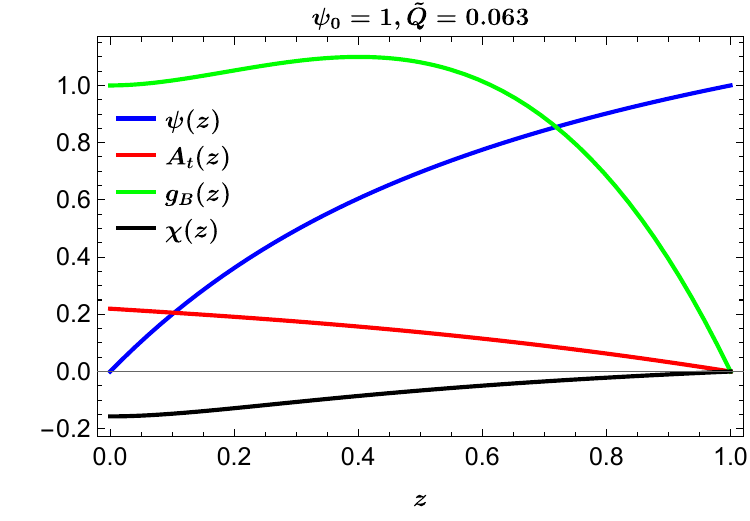}
    \includegraphics[height=0.31\linewidth]{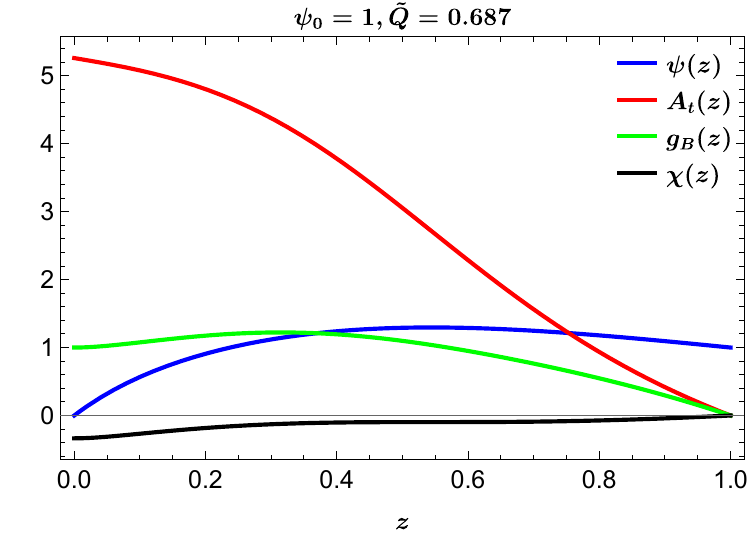}
    \caption{Example profiles of field functions for tachyonic-hairy black hole solutions (left) and scalar-hairy black hole solutions (right).}
    \label{fig=profile_nega}
\end{figure}

Let us first identify the two distinct types of hairy black holes appearing in our numerical solutions.
Representative profiles of the field functions are displayed in Fig.~\ref{fig=profile_nega}.
For small electric charge, the scalar field exhibits a monotonic decay, indicating that the scalar potential term dominates the formation of the black hole’s scalar hair.
In contrast, as the electric charge increases, the scalar field develops a nonmonotonic behavior: it grows along the radial direction near the horizon, reaches a local maximum, and then decays toward zero at spatial infinity.
When $\psi_0$ is fixed, the metric functions $g_B(z)$ and $\chi(z)$ show no qualitative difference between the two black hole solutions, whereas the Maxwell functions display significant quantitative deviations.
Once the scalar hair is allowed to vary freely, it becomes clear that none of the three aforementioned field functions alone can effectively distinguish the two types of hairy black holes.
The most prominent difference between them lies precisely in the behavior of the scalar field near the horizon, specifically, in its growth and decay pattern in that region.

\begin{figure}[thbp]
    \centering
    \includegraphics[height=0.31\linewidth]{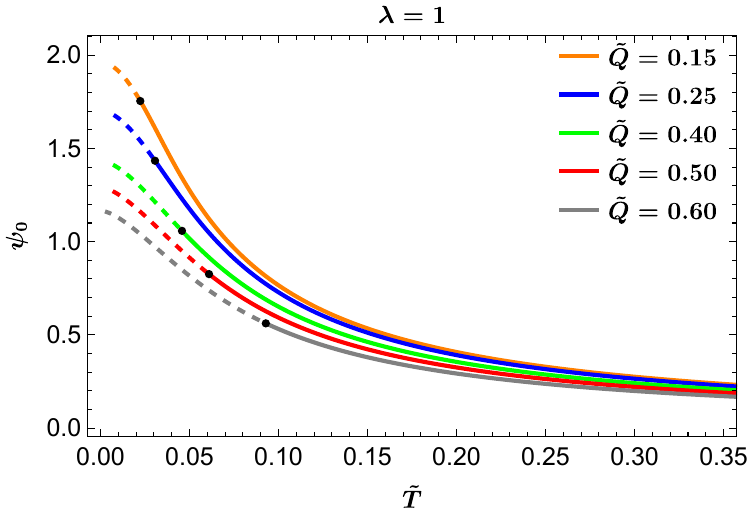}
    \includegraphics[height=0.31\linewidth]{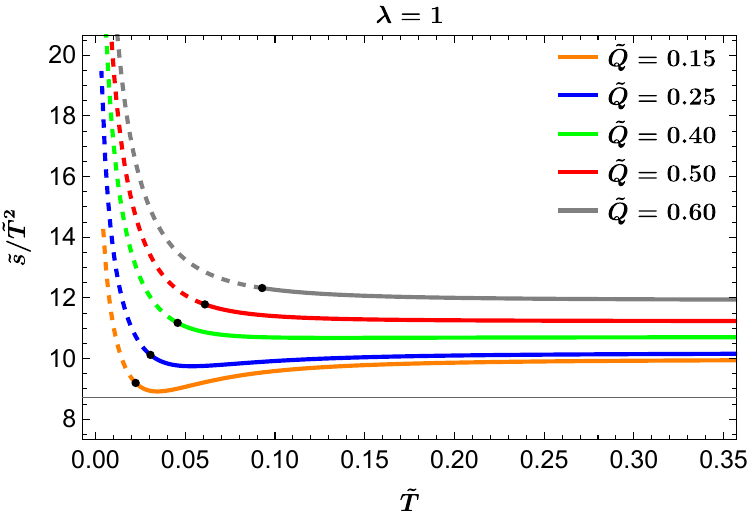}
    \caption{Scalar hair (left) and induced entropy density (right) as functions of temperature for several electric charges. Dashed segments denote scalar-hairy solutions, while solid segments denote tachyonic-hairy solutions, the black dots mark the phase boundaries between them.}
    \label{fig=hair}
\end{figure}

This observation motivates an investigation of the phase boundary between the two hairy black hole branches under the extremal condition $\psi'(1)=0$. 
To this end, we study the scalar hair $\psi_0$ as a function of temperature at fixed electric charge $\tilde{Q}$.
The left panel of Fig.~\ref{fig=hair} shows a smooth transition from tachyonic-hairy to scalar-hairy solutions as the temperature is lowered. 
At a given $\tilde{Q}$, the scalar-hairy phase consistently carries substantially more scalar hair than the tachyonic-hairy phase. 
The phase boundary, indicated by the black dot in the figure, shifts to higher temperature but to smaller values of $\psi_0$ as the electric charge is increased. 
These trends suggest possible phase transitions between the two hairy phases in our numerical solutions.
The induced entropy density $\tilde{s}/\tilde{T}^2$, shown in the right panel of Fig.~\ref{fig=hair}, also varies smoothly across the transition. 
We find that scalar-hairy black holes have larger entropy than tachyonic-hairy ones, suggesting that the scalar-hairy branch is thermodynamically preferred. 
Notably, at low electric charge, the entropy density exhibits nonmonotonic behavior: it decreases before the phase boundary and then increases after passing a local minimum. 
This feature motivates further study of the phase structure near the low-temperature region of the phase boundary.

\begin{figure}[thbp]
    \centering
    \includegraphics[height=0.31\linewidth]{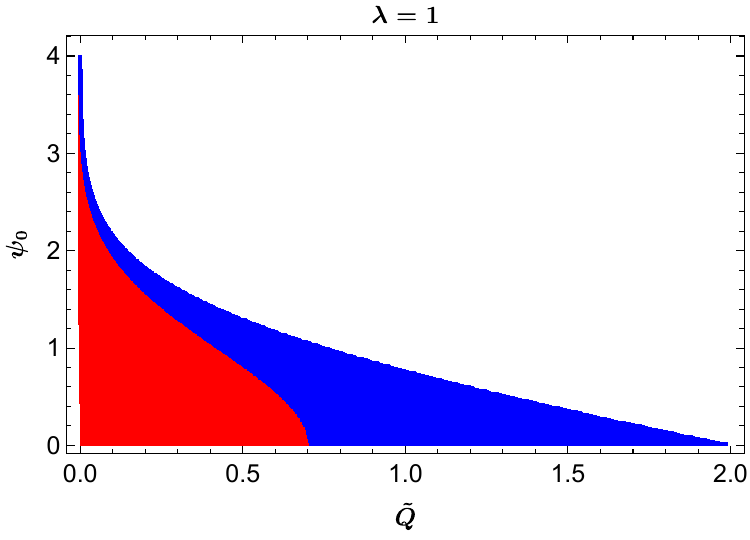}
    \includegraphics[height=0.31\linewidth]{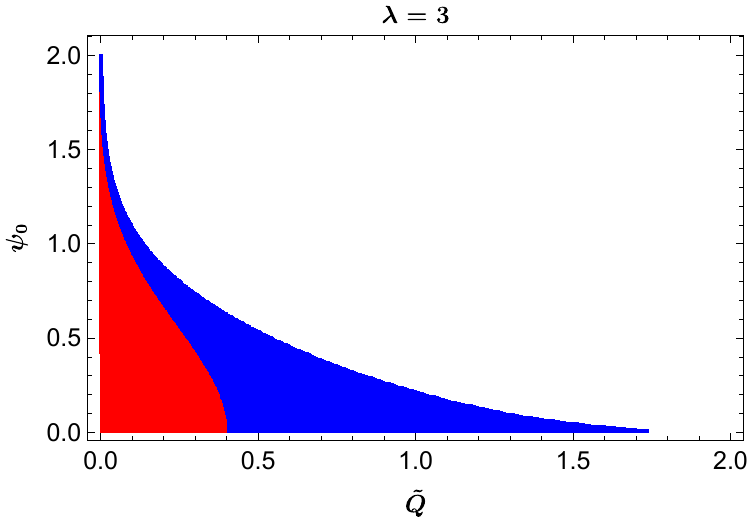}
    \caption{Existence domains of the two distinct hairy black hole solutions in the parameter space of scalar hair $\psi_0$ and electric charge $\tilde{Q}$ for $\lambda=1$ and $\lambda=3$, respectively. The tachyonic-hairy phase (red region) appears at small $\tilde{Q}$, while the scalar-hairy phase (blue region) dominates at large $\tilde{Q}$.}
    \label{fig=Qdomain}
\end{figure}

We now examine the existence domains of the hairy black hole solutions in the $(\tilde{Q},\psi_0)$ parameter space, as shown in Fig.~\ref{fig=Qdomain}.
Consistent with the previous discussion, the tachyonic-hairy phase appears in the small-$\tilde{Q}$ region, while the scalar-hairy phase dominates at larger $\tilde{Q}$.
For $\lambda=1$, the allowed range of $\tilde{Q}$ for hairy solutions extends up to the RN–AdS limit $\tilde{Q}_{\text{max}}=2$ as $\psi_0 \to 0$.
In particular, the existence ranges of $\tilde{Q}$ for both hairy phases shrink as the scalar hair $\psi_0$ increases, eventually vanishing at the maximum value of $\psi_0$. 
Increasing the coupling parameter $\lambda$ reduces this maximum value of $\psi_0$ for both the tachyonic-hairy and scalar-hairy branches. 
Moreover, the range of electric charge in which the tachyonic-hairy solutions exist is further suppressed by stronger coupling, whereas the scalar-hairy solutions remain largely unaffected. 
This behavior indicates that the coupling strength $\lambda$ governs the competition between the two types of hairy black hole solutions.

\subsection{Phase diagram}\label{sec=negamass_phase}

\begin{figure}[thbp]
    \centering
    \includegraphics[height=0.31\linewidth]{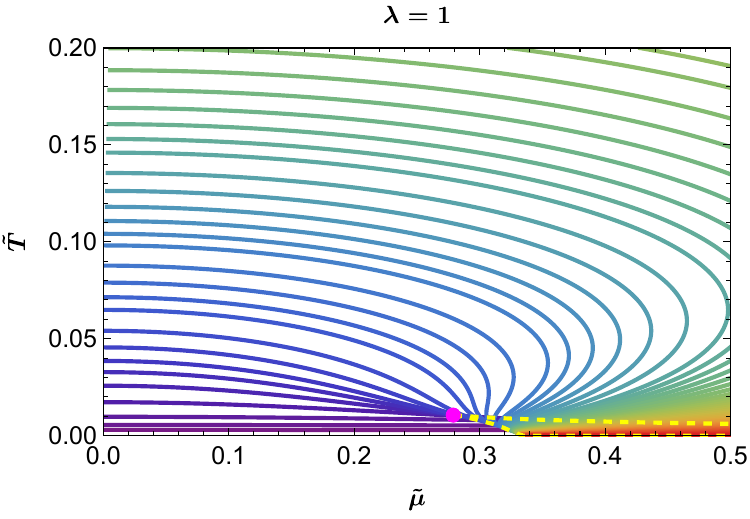}
    \includegraphics[height=0.31\linewidth]{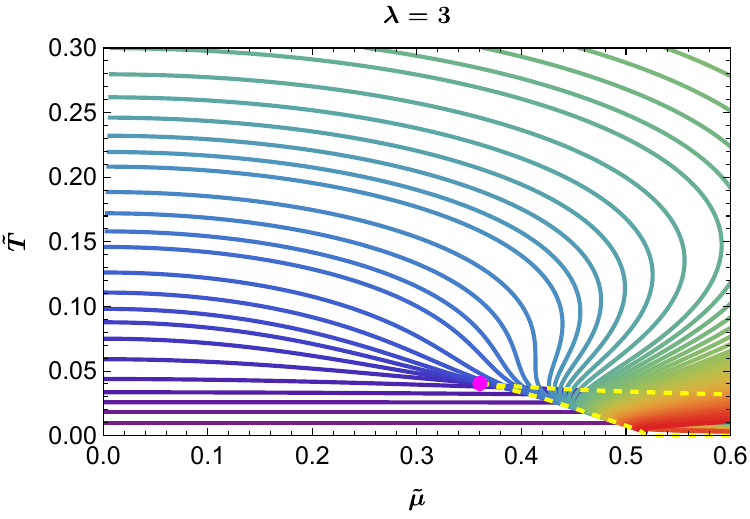}
    \caption{Hairy black hole solutions for different $\psi_0$ in the parameter space of $(\tilde{\mu},\tilde{T})$ with several coupling value. The yellow dashed lines indicate the boundaries of the overlap region, while magenta dots mark their start point.}
    \label{fig=phase_negamass}
\end{figure}

Now we focus on the phase diagram of the boundary field theory. 
In the parameter space of $(\tilde{\mu},\tilde{T})$, as shown in Fig.~\ref{fig=phase_negamass}, we show the numerical solutions with the relation between the temperature and chemical potential for different $\psi_0$.
The results show rich structure, notably, an overlap region appears at low temperature and relatively large chemical potential.
We see that the temperature curves in the low-temperature region generally exhibit a monotonic relationship with chemical potential. 
However, as the temperature increases, multiple temperature curves converge to a common point and gradually develop nonmonotonic behavior. 
In the high-temperature region, these curves cluster near the convergence point at high chemical potentials and merge with the low-temperature curves, forming a regular overlap region that begins at this convergence point.
The boundary of this overlapping region is indicated by yellow dashed lines in the figure, while its start point is marked by a magenta dot.
As $\lambda$ increases, the overlap region enlarges and its endpoint moves to higher temperature and higher chemical potential region.

\begin{figure}[thbp]
    \centering
    \includegraphics[height=0.31\linewidth]{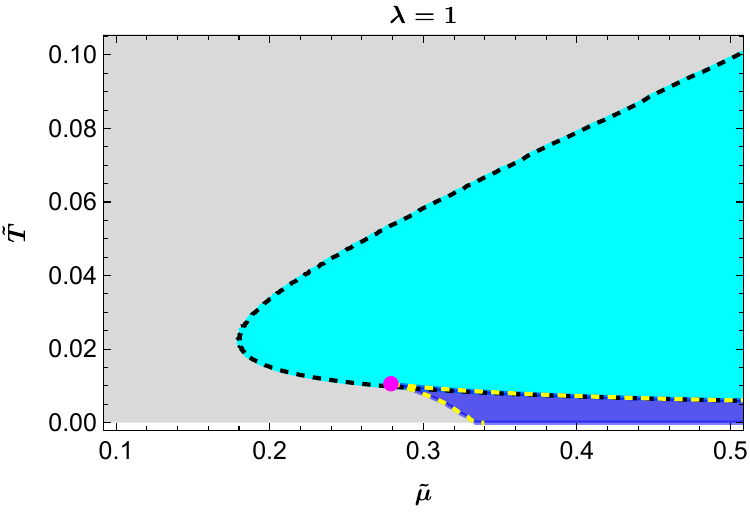}
    \includegraphics[height=0.31\linewidth]{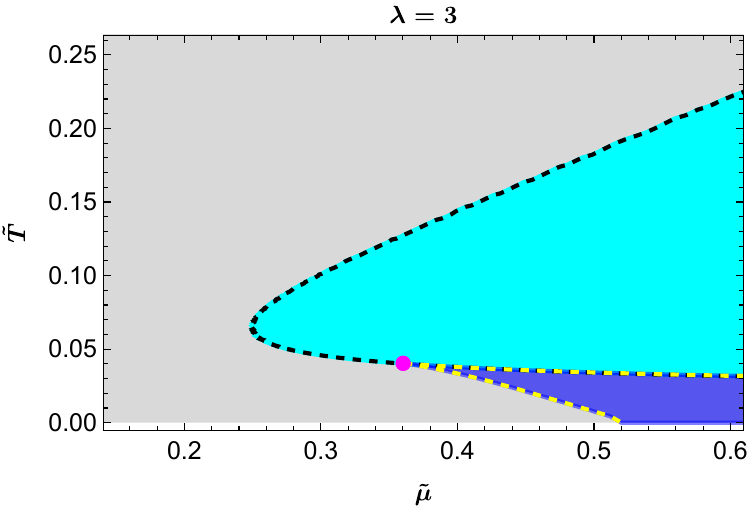}
    \caption{Existence domain of the two distinct hairy phases in the parameter space of $(\tilde{\mu},\tilde{T})$ for several values of coupling parameter. The gray region represents the tachyonic-hairy phase, while cyan region denotes the scalar-hairy phase. Their phase boundary is indicated by the black dashed curves. The blue region corresponds to the overlap region, bounded by yellow dashed lines, whose start point is marked by a magenta dot.}
    \label{fig=phase_diagm}
\end{figure}

\begin{figure}[thbp]
    \centering
    \includegraphics[height=0.26\linewidth]{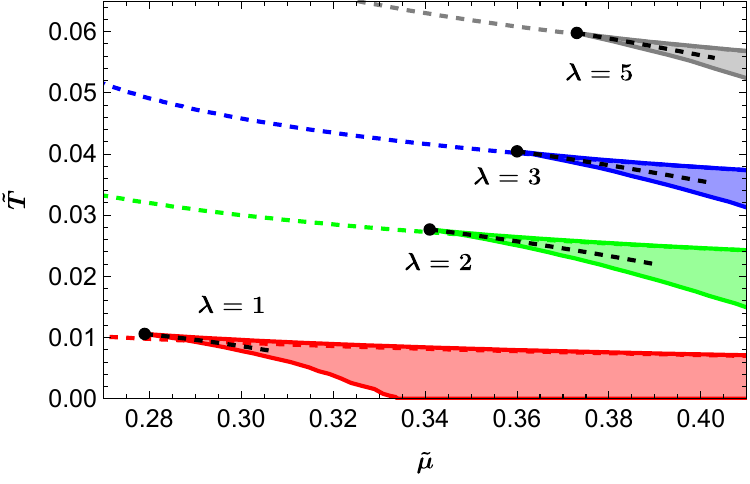}
    \includegraphics[height=0.26\linewidth]{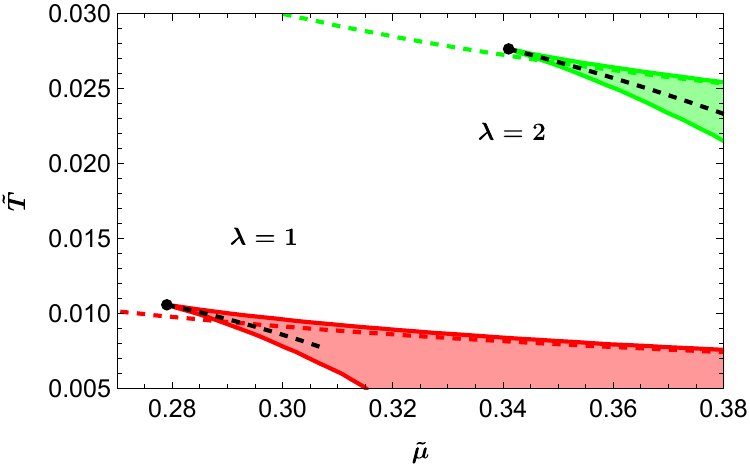}
    \caption{First-order phase transition lines (black dashed lines) for different values of coupling parameter, with critical point marked by the black dot. Colored dashed lines indicate the phase boundary of the two hairy solutions, while the colored regions bounded by colored solid lines denote the overlap regions.}
    \label{fig=phasediagall}
\end{figure}

Although the foregoing analysis strongly suggests a phase transition in the overlap region, it does not provide definitive evidence or a concrete physical picture of the phase structure. 
Our detailed study of the hairy black hole solutions fills this gap. 
Using the distinctions identified for the tachyonic-hairy and scalar-hairy branches, we map the regions occupied by each phase. 
In Fig.~\ref{fig=phase_diagm}, the gray region denotes the tachyonic-hairy phase and the cyan region denotes the scalar-hairy phase. 
Both terminate at a phase boundary described by the black dashed curve. 
The blue region indicates the overlap region, whose boundary is plotted with yellow dashed lines and which terminates at a start point marked by a magenta dot, as illustrated in Fig.~\ref{fig=phase_negamass}.
A closer inspection shows that the overlap region nearly coincides with the coexistence region of the two phases, with only minor differences. 
For weak coupling, the yellow dashed boundary of the overlap region does not exactly coincide with the black dashed phase boundary.
Consequently, the overlap start point lies slightly above the phase boundary (this is visible in the right panel of Fig.~\ref{fig=phasediagall}). 
Thus, for small coupling the true coexistence region corresponds to the portion of the blue area lying below the black dashed line. 
By contrast, for larger coupling values, the yellow and black dashed lines coincide within numerical accuracy, and the overlap (blue) region then precisely corresponds to the coexistence region.

Consistent with the preceding discussion, the coupling parameter controls the competition between the two distinct hairy solutions. 
Eq.~\eqref{eq=Ch1} shows that $\lambda$ regulates the balance between the coupling function and the scalar mass term. 
From Fig.~\ref{fig=phase_negamass}, we observe that the domain of the scalar-hairy phase including its coexistence portion, expands as $\lambda$ increases, indicating that the scalar-hairy branch becomes increasingly favored and occupies a larger region of parameter space. 
We also find that the start point of the overlap region shifts to higher temperature and larger chemical potential as $\lambda$ grows.

\begin{figure}[thbp]
    \centering
    \includegraphics[height=0.26\linewidth]{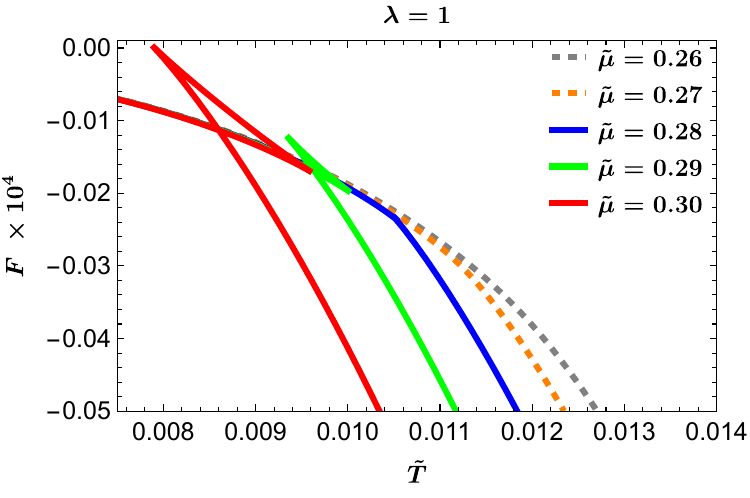}
    \includegraphics[height=0.26\linewidth]{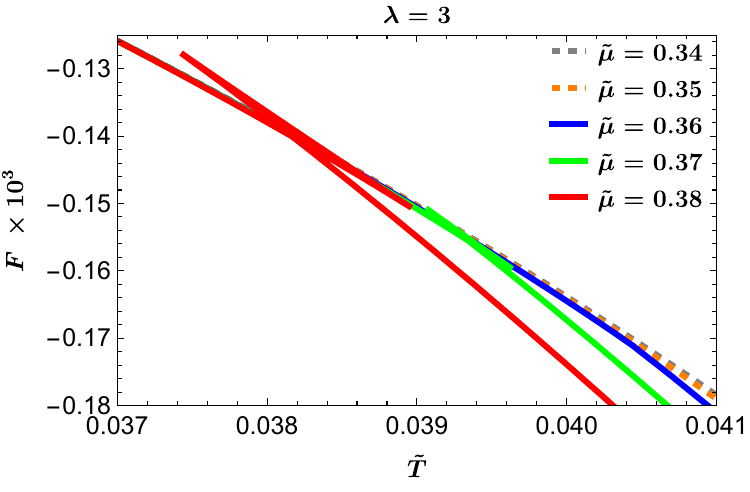}
    \includegraphics[height=0.26\linewidth]{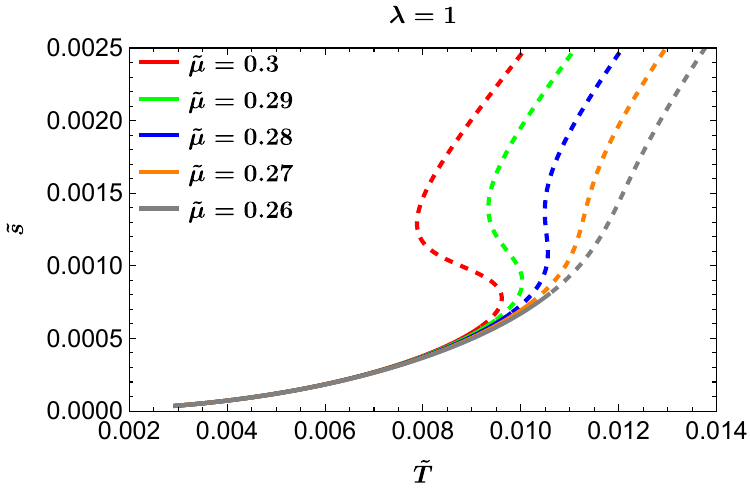}
    \includegraphics[height=0.26\linewidth]{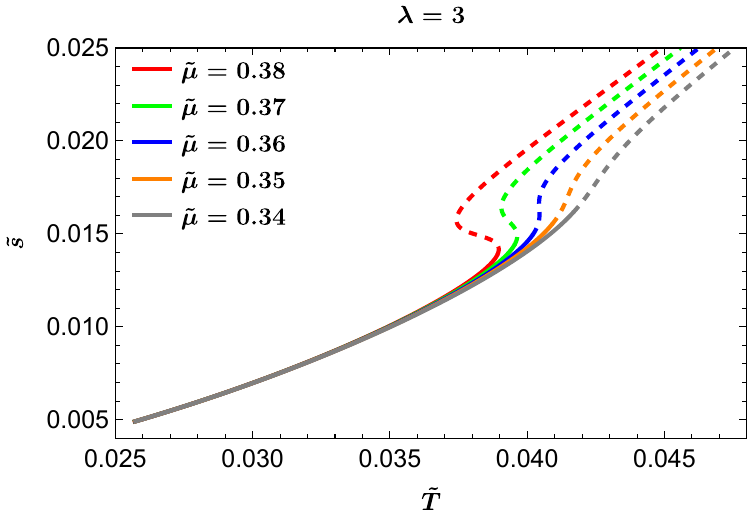}
    \caption{Top: Free energy density as a function of the temperature for several chemical potentials crossing the lower phase boundary for $\lambda=1$ and $\lambda=3$, respectively; Bottom: Corresponding entropy density as a function of the temperature, with solid segments representing tachyonic-hairy phases and dashed segments denoting the scalar-hairy phases.}
    \label{fig=overlap_phase}
\end{figure}

Therefore, any potential phase transition is expected to occur near the phase boundary. 
From previous experience, along the segment of the phase boundary above the start point where no phase coexistence is present, the transition between the two phases is smooth and gradual, making a genuine phase transition unlikely. 
In contrast, below the start point, the coexistence of the two phases indicates that the change between them is extremely sharp, and the actual location of the phase transition may not lie exactly on either the phase boundary or the edge of the overlap region.
To identify a possible phase transition along the low-temperature boundary, particularly within the coexistence region, we compute the entropy density and the corresponding free energy density, with $dF=\tilde{s}d\tilde{T}$ at fixed chemical potential $\tilde{\mu}$. 
The results are presented in Fig.~\ref{fig=overlap_phase}, which illustrates the behavior as temperature increases across the lower part of the phase boundary. 
As indicated by the dashed curves, the free energy density decreases monotonically with temperature when $\tilde{\mu}$ is small. 
However, once the chemical potential exceeds a critical value, the free energy density develops a characteristic swallow-tail structure as a function of temperature. 
Correspondingly, the entropy density follows a zigzag trajectory instead of a simple monotonic increase.
This behavior implies that the thermodynamic process traversing the overlap region exhibits multivalued features, confirming the existence of a first-order phase transition near the low-temperature boundary. 
An important observation from the entropy density behavior in the lower panels of Fig.~\ref{fig=overlap_phase} is that the first-order phase transition originates not from the coexistence of the two phases, but from the self-overlapping structure within the scalar-hairy phase itself. 
A representative example is shown by the blue curve in the bottom-left panel. 
Despite the absence of coexistence between the solid and dashed segments, the zigzag path of the dashed segment still signals a first-order transition. 
This result indicates that the critical chemical potential associated with the phase transition is located at the start point of the overlap region, suggesting that the overlap region, rather than the coexistence region, is more directly tied to the occurrence of the phase transition.

The critical temperature of the first-order phase transition is determined by the intersection point of the swallow-tail structure in the free energy density, allowing us to trace the precise transition line. 
The results are shown in Fig.~\ref{fig=phasediagall}, where the black dashed lines represent first-order transition lines for different coupling parameters $\lambda$. 
These lines originate at the critical points located at the start points of the overlap regions (colored areas) and extend into these regions as the chemical potential increases. 
The right panel provides a zoomed view of the left panel, showing that for small $\lambda$, the critical point initially lies above the phase boundary (colored dashed line) and crosses it as the transition progresses to higher chemical potentials. 
For larger $\lambda$, this complication disappears, since the overlap start points coincide with the phase boundary, and the transition lines extend directly inward into the overlap region. 

\begin{figure}[thbp]
    \centering
    \includegraphics[height=0.26\linewidth]{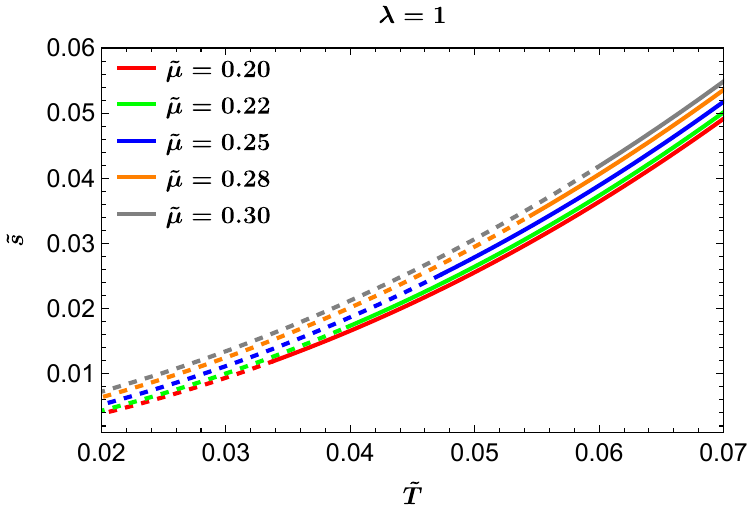}
    \includegraphics[height=0.26\linewidth]{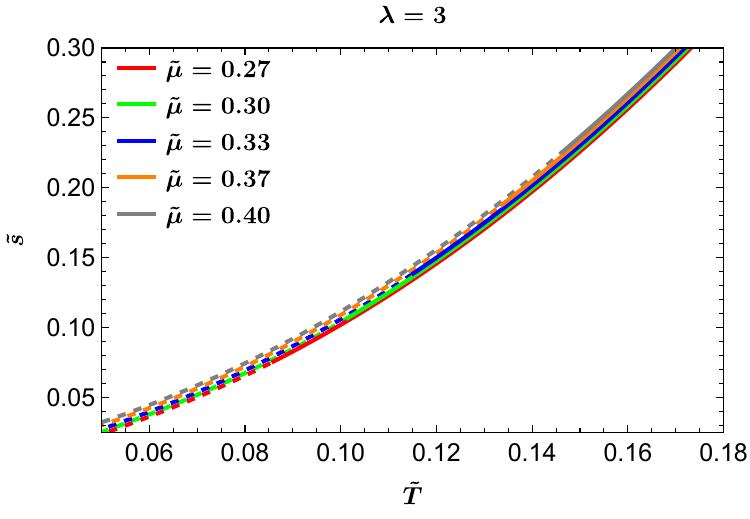}
    \includegraphics[height=0.26\linewidth]{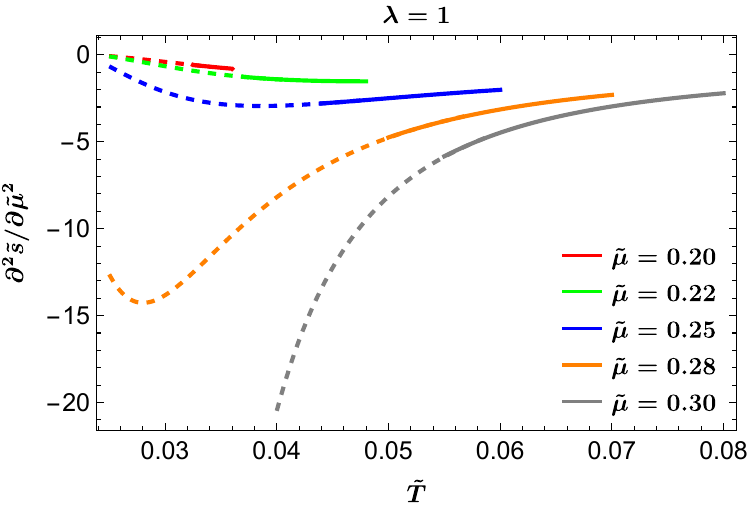}
    \includegraphics[height=0.26\linewidth]{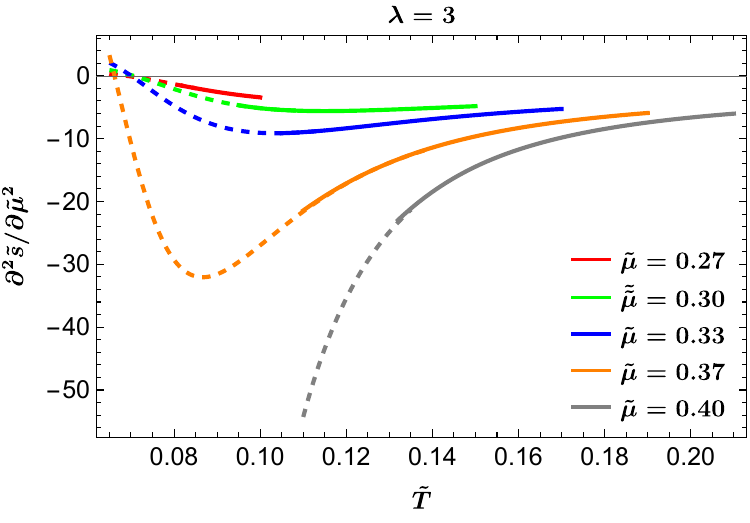}
    \caption{Top: Free energy density as a function of the temperature for several chemical potentials acrossing the high-temperature phase boundary for $\lambda=1$ and $\lambda=3$, respectively; Bottom: Corresponding second derivative of entropy density with respect to the chemical potential, with solid segments representing tachyonic-hairy phases and dashed segments denoting the scalar-hairy phases.}
    \label{fig=phasebdy_high}
\end{figure}

We also examine the potential for phase transitions along the high-temperature phase boundary outside the overlap region. 
As shown in Fig.~\ref{fig=phasebdy_high}, the entropy density and its second derivative with respect to chemical potential vary smoothly across this boundary for $\lambda=1$ and $\lambda=3$, indicating no phase transition up to at least third-order. 
Together with the observation that the chemical potential is below the critical value, this confirms that no phase transition occurs along the boundary above the start point.
In conclusion, Fig.~\ref{fig=phasediagall} presents the final phase diagram for the boundary field theory. 
It indicates a first-order phase transition between the tachyonic-hairy and scalar-hairy phases, beginning at a critical point located at the start point of the overlap region shown in Fig.~\ref{fig=phase_negamass}.

\section{Conclusion and discussion}

Unlike spontaneous scalarization, the introduction of a positive coupling parameter in the EMS theory can also induce hairy black holes in flat spacetime. 
In AdS spacetime, we consider a real massive scalar field within the EMS model and examine both scalar-hairy black hole solutions and tachyonic-hairy solutions driven by the scalar potential. 
When the scalar potential vanishes, scalar-hairy black holes emerge with profiles and properties similar to those observed in flat spacetime.
The presence of the scalar potential additionally induces tachyonic-hairy solutions, leading to the coexistence of these two distinct hairy phases in different regions of the parameter space. 
The phase diagram reveals a first-order phase transition line between the tachyonic-hairy and scalar-hairy phases, originating at a critical point in the extreme temperature and chemical potential regime. 
Our detailed analysis shows that this phase transition is directly associated with the self-overlap region of the scalar-hairy phase and its start point. 
In the case of small coupling, the coexistence region of tachyonic- and scalar-hairy phases forms only part of the overlap region, and the critical point of the phase transition is located at the start point of the overlap, which lies above the phase boundary. 
This complex situation becomes much simple in the large coupling case, where the coexistence region coincides with the overlap region, and the critical point lies directly on the phase boundary.

It is important to note that this work provides a systematic and comprehensive clarification of the previously ambiguous discussions in~\cite{Guo:2025xwh,Guo:2024ymo,Guo:2025rxh}. 
In~\cite{Guo:2024ymo,Guo:2025rxh}, two types of hairy black hole solutions arising from spontaneous scalarization were proposed, and first-order phase transitions in the boundary field theories of both simplified and improved EMD models were discussed. 
In~\cite{Guo:2025xwh}, it was shown that in asymptotically flat spacetime, EMS theory with negative coupling leads to spontaneous scalarization, whereas positive coupling does not induce tachyonic instability. 
Consequently, novel scalar-hairy black holes with predominantly monotonic growth were observed.
Our current results are qualitatively consistent with these earlier studies and allow for a precise determination of the first-order phase transition line and its critical point. 
Notably, scalar field growth behavior is observed in all cases, supporting the conclusion that scalar-hairy black holes play a central role in first-order phase transitions within holographic EMD models. 
Furthermore, adjusting the coupling parameter systematically shifts both the phase transition line and its critical point within the phase diagram. 
Larger coupling moves the critical point to higher temperatures and chemical potentials, suggesting that QCD-like phase transitions can be simulated not only via methods where scalar field sources set energy scales~\cite{Cai:2022omk,Critelli:2017oub,Grefa:2021qvt}, but also by tuning the coupling strength, which effectively adjusts the relevant energy scales.

One remaining issue is that, while the current work differs from~\cite{Guo:2024ymo} in the form of the scalar potential, both models agree at leading order under the weak-field approximation. 
However, the third-order phase transition reported in~\cite{Guo:2024ymo} is not observed here, suggesting that the underlying mechanism of phase transitions for hairy black holes remains incompletely understood and warrants further investigation.

\section*{Acknowledgments}
We appreciate Miok Park for helpful correspondence.
This work is supported by the Institute for Basic Science (Grant No. IBS-R018-Y1). 
H. L is also supported by National Natural Science Foundation of China under Grant No.12305071.
Y.S.M. is also supported by the National Research Foundation of Korea (NRF) grant funded by the Korea government(MSIT) (RS-2022-NR069013).

\newpage

\bibliographystyle{JHEP}
\bibliography{refs}

\end{document}